\documentclass[aps]{revtex4}
\usepackage{amsbsy}
\usepackage{graphicx,color}
\usepackage{amsfonts}
\usepackage{amsmath}
\usepackage{amssymb}

\newcommand{\ds}{ _{\downarrow}}
\newcommand{\us}{ _{\uparrow}}
\newcommand{\up}{\uparrow}
\newcommand{\down}{\downarrow}

\begin{document}
\title{Atom-dimer scattering length for fermions with different masses: analytical study of limiting cases}
\author{F. Alzetto$^{(a)}$, R. Combescot$^{(a),(b)}$ and  X. Leyronas$^{(a)}$}
\address{(a) Laboratoire de Physique Statistique, Ecole Normale Sup\'erieure, UPMC  
Paris 06, Universit\'e Paris Diderot, CNRS, 24 rue Lhomond, 75005 Paris,  
France.}
\address{(b) Institut Universitaire de France}
\date{Received \today}
\pacs{PACS numbers : 03.75.Kk,  05.30.-d, 47.37.+q, 67.90.+z }

\begin{abstract}
We consider the problem of obtaining the scattering length for a fermion colliding with a dimer, formed from a fermion identical
to the incident one and another different fermion. This is done in the universal regime where the range of interactions
is short enough so that the scattering length $a$ for non identical fermions is the only relevant quantity. This is the 
generalization to fermions with different masses of the problem solved long ago by Skorniakov and Ter-Martirosian
for particles with equal masses. We solve this problem analytically in the two limiting cases where the mass of the solitary 
fermion is very large or very small compared to the mass of the two other identical fermions. This is done both for the value of
the scattering length and for the function entering the Skorniakov-Ter-Martirosian integral equation, for which simple
explicit expressions are obtained. 
\end{abstract}
\maketitle
\section{INTRODUCTION}

Recent years have seen very important developments in the field of ultracold fermionic gases \cite{gps}.
In addition to its own intrinsic interest, this field has a strong overlap with problems arising not only in condensed matter physics,
where the degenerate electron gas is an essential ingredient of metal physics, but also in nuclear physics and in quark matter
physics \cite{casal}. A major strong point of these cold gases systems is the extreme simplicity of the effective interaction
between fermions of different species, which can be fully described by the mere knowledge of the scattering length $a$, while
there is no effective interaction between same species fermions, since this has to occur in the s-wave channel because of the
very low temperature and energies involved and this s-wave scattering is forbidden by Pauli principle (we assume that
no p-wave or higher-wave resonance occurs). In addition the scattering
length can be changed, in the vicinity of Feshbach resonances, by merely modifying the static uniform magnetic field applied
to the trapped gas. This offers the fascinating possibility to control at will the strength of the interaction.

The first explorations of these gases have dealt with systems containing two different kinds of fermions with same particle number and  same 
mass, in practice fermionic atoms belonging to the same element but being in different hyperfine states. This is the most convenient
and the most natural system to explore experimentally, and in addition the physical situation is the same as for electron in metals
which may have their spins up or down and are roughly in equal number. A fair exploration has been made of the normal state
and mostly of the superfluid state where the whole BEC-BCS crossover has been studied to a large extent.
More recently the research activity has moved toward 
systems where there is a sizeable imbalance between the two fermionic populations \cite{gps} and where a great deal of interesting
physics has been found. Quite recently some attention for fermions with different masses has arisen. This is because new interesting 
physics, both for normal and superfluid phases, is expected when the mass ratio becomes large \cite{pao,parish,blume,baranov,ops,brs, diener}. 
Experimental developments in this direction are in progress in several 
laboratories in the world \cite{munich,innsb} and degeneracy has already been reached in these mixtures. Similarly several quite recent theoretical
works have focused on this situation \cite{ggsc, bdgc}, with an emphasis on systems with a small number of particles where multiparticle bound 
states are expected to occur.

As we have mentioned the most interesting situations occur when there is a s-wave Feshbach resonance, which is directly linked to the
appearance of a bound state, or molecular state, or dimer, between two atoms belonging to different fermion species. This will
occur on the so-called BEC side of the resonance, and on this side these dimers play clearly an essential role in the physics
of the gas. As it is quite common we call the two involved fermion species $\up$ and $\down$ spin, by analogy with electrons in solids, 
even if we are actually interested in fermions belonging to different atomic elements. If the masses of these atoms are respectively
$m\us$ and $m\ds$, the mass of the dimer is the total mass $M=m\us + m\ds$. The binding energy is $E_b=1/(2\mu a^2)$
where $\mu =m\us m\ds /(m\us + m\ds)$ is the reduced mass. In addition to these dimers we have naturally in general on the BEC side
isolated $m\us$ and $m\ds$ fermionic atoms. This will occur for example at zero temperature if there is an imbalance between the
two fermionic populations so that all fermions can not be paired into dimers. Or this may also occur at non zero temperature because
the dimers will be thermally broken. Hence scattering between dimer and isolated atoms will happen, and in cold gases this scattering 
will be fully characterized by the atom-dimer scattering length $a_3$. Clearly this is, beyond the up-down scattering length $a$, 
one of the basic quantities coming in the physics of these systems. This is the quantity we are interested in. Quite specifically we will
investigate the scattering length $a_3$ of a $\up$ atom by a dimer.

Naturally an important feature in these kind of systems is their stability \cite{gps}. It is known that inelastic three-body collisions are a 
major process for loss rate in these systems, being responsible for their limited experimental lifetime. Obviously the existence of dimers
favours  these processes since two atoms are already fairly nearby in these bound states.
This is indeed known to be a major problem in bosonic systems. Fortunately in fermionic systems, with which we are concerned here, these
processes are comparatively restricted by the fact that, qualitatively, Pauli exclusion keeps the same species fermions apart, limiting the
overlap with deep bound states responsible for relaxation. This is origin of the fairly long lifetime found for fermionic systems made of the same
atomic element. However the situation might be somewhat different for fermions with different masses and this is obviously a question
under current experimental investigation. Moreover in these systems dimer-dimer collisions provide also an important relaxation channel
\cite{gps}. Finally \cite{bdgc} three-body and four-body resonances occuring in these systems might lead to efficient additional loss channels.
In the present paper we will assume that, nevertheless, the stability of the system is high enough for our investigation to be meaningful.

The calculation of $a_3$ is a long standing problem and, for equal masses, the problem has been solved long ago by Skorniakov and 
Ter-Martirosian \cite{stm}. The generalization to different masses is naturally quite easy, either following the original paper \cite{stm} 
or making use of our recent work \cite{bkkcl}, as it has been done recently by Iskin and Sa de Melo \cite{ism}. The resulting equation 
has been quite recently fully investigated numerically by Iskin \cite{isk}. Another approach has been used by Petrov \cite{petr} who
has obtained $a_3$ by solving the 3-body Schr\"odinger equation with the appropriate boundary conditions imposed by the scattering
length $a$.

While the above works solve completely the problem as far as numbers are concerned, we feel that a deeper understanding of the
involved equations is warranted for such a basic problem. Moreover such a full control provides a first step to address more complicated
problems which will arise in many-body problems, involving the same building blocks, which can be formulated with the same
Green's functions formalism, using the same Feynman diagrams as we have done for example in \cite{lhy}. Such a control
means in principle a complete analytical investigation. Unfortunately this does not seem possible in the general case. On the other hand
we have been able to carry out this investigation in the limiting cases where the mass ratio $r=m\ds / m\us$ is either very large or
very small and the purpose of this paper is to present our results. We note that for the small $r$ limit Petrov \cite{petr} has done
briefly a similar investigation based on a Born-Oppenheimer approximation within the Schr\"odinger equation. 
Here we address this problem within the Skorniakov and Ter-Martirosian approach, in a more systematic way and
with more specific and precise results. After presenting the basic equation in section \ref{base}, we consider in section \ref{vhma} the
case where the mass ratio $r$ is very large, and then in section \ref{vlm} the case where it is very small. Since this last case is
more involved section \ref{vlm} contains several subsections. Let us finally mention that an important ingredient in all the few-body problems
is the existence of bound states of the whole system under study, in our case bound states of the $\up$ atom and the dimer. This
has been studied in great details quite recently by Kartavtsev and Malykh \cite{karma} and these bound states include in particular the Efimov states
\cite{efimov} which have received recently a great deal of attention. However since we are here only interested in fermions and
in s-wave states, these bound states do not come in our problem since they appear only for non-zero angular momentum \cite{karma}.

\section{Basic equations}\label{base}

Our basic equation is obtained by writing directly \cite{bkkcl} the integral equation for the full dimer-fermion scattering vertex $T_3(p_1,p_2;P)$, 
where $P$ is the (conserved) momentum-energy of the particles $P \equiv \{\textbf{P},E\}$, with
$\textbf{P}$ the total momentum and $E$  the total frequency of incoming particles. $p_1$ is the momentum-energy of the incoming
$\up$ fermion and $p_2$ its outgoing value.

The scattering length is obtained by considering the case of total momentum $\textbf{P}={\bf 0}$ and total energy $E=-E_b$.
The incoming and outgoing momentum-energy of the scattering fermion should also be zero $p_1=p_2=0$. Introducing the
"on-the-shell" value of $T_3$ for the incoming fermion and the related function $a_3(\mathbf{p})$ according to:
\begin{equation}
    \left(\frac{2\pi}{\mu ^2a}\right)T_3\left(\{\mathbf{p},p^2/2m\us\},0;
    \{{\bf 0},-E_b\right\}) \left[1+\sqrt{1+\frac{\mu }{\mu _T}(pa)^2}\,\right]=2\,\frac{2\pi}{\mu _T}\,a_3(\mathbf{p}).
\end{equation}
the atom-dimer scattering length $a_3$ is merely given by $a_3=a_3(\mathbf{0})$. Here $\mu _T$ is the atom-dimer reduced mass
$\mu _T = m\us M /(m\us + M)=m\us (m\us + m\ds)/(2 m\us +m\ds)$. Then the integral equation for $T_3$, 
restricted to "on-the-shell" values, becomes the generalization
for different masses of the equation of Skorniakov and Ter-Martirosian for $a_3(\mathbf{p})$, namely:
\begin{equation}\label{a_3}
    \frac{\mu}{\mu _T} 
   \;\frac{\bar{a}_3(\mathbf{\bar{p}})}{1+\sqrt{1+ \frac{\mu}{\mu _T}\,\bar{p}^2}}
    =\frac{1}{1+\bar{p}^2}-4\pi\int\frac{d^{3}\mathbf{\bar{q}}}{(2\pi)^3}\;\frac{\bar{a}_3(\mathbf{\bar{q}})}
    {\bar{q}^2}\,\frac{1}{1+\bar{p}^2+\bar{q}^2+2\, \frac{\mu}{m\ds}\,\mathbf{\bar{p} \cdot \bar{q}}}
\end{equation}
where we have used reduced units by setting $a_3/a=\bar{a}_3$ and ${\bf p}=\mathbf{\bar{p}}/a$ 
and ${\bf q}=\mathbf{\bar{q}}/a$.
The angular integration is easily performed, since $\bar{a}_3(\mathbf{\bar{p}})$ depends clearly only on $|\mathbf{\bar{p}}|$. Using the simpler
notations $R=\mu /\mu _T$, $R'=2 \mu /m\ds$ (they are related by $R+R'^2/4=1$) , together with $x=|\mathbf{\bar{p}}|=\bar{p}$, 
$y=|\mathbf{\bar{q}}|=\bar{q}$ and $f(x)=\bar{a}_3(\mathbf{\bar{p}})$, one finds:
\begin{equation}\label{eqf}
R\;\frac{f(x)}{1+\sqrt{1+R x^2}}
    =\frac{1}{1+x^2}-
\frac{1}{\pi R'} \int_{0}^{\infty} dy \;\frac{f(y)}
    {xy}\,\ln \frac{1+x^2+y^2+R'xy}{1+x^2+y^2-R'xy}
\end{equation}
For equal masses, $R=3/4$ and $R'= 1$ and solving this equation gives the well-known result \cite{stm} for the
atom-dimer scattering length $a_3 = 1.18 \,a$.

The generalization of the Skorniakov and Ter-Martirosian equation following the above lines has been obtained by Iskin and S\'a de Melo \cite{ism},
who have solved the equation numerically and plotted $a_3/a$ as a function of the mass ratio $r=m\ds / m\us$. 
This has been taken up in great details by Iskin \cite{isk}.
Let us recall that the homogeneous equation corresponding to Eq.(\ref{eqf}) has no solution, corresponding to the
physical fact that \cite{karma,efimov} there are no Efimov states, nor any other bound state, for the three fermions problem we are considering.

Before turning to the limiting cases we are interested in, we present first some general results which will be useful in the following.
Let us first consider the behaviour of $f(x)$ when $x\rightarrow \infty$. If one assumes (which is confirmed numerically)
that, in the integral in the right-hand side of Eq.(\ref{eqf}), $f(y)$ goes rapidly enough to zero for large $y$ (possibly with an oscillating behaviour) so that the integral converges, one can consider $y$ as effectively bounded, and for
$x \rightarrow \infty$ expand the logarithm. This leads from Eq.(\ref{eqf}), in this large $x$ range, to the simplified equation:
\begin{equation}\label{eqfasym}
\sqrt{R}\;\frac{f(x)}{x}
    =\frac{1}{x^2}\;\left[1\;-
\frac{2}{\pi} \int_{0}^{\infty} dy \;f(y)\;\right]
\end{equation}
Apparently this leads to $f(x) \sim 1/x$, in contradiction with our starting hypothesis since, in this case, the integral
in right-hand side of Eq.(\ref{eqfasym}) diverges. The only possible escape is that the coefficient of $1/x^2$ in this
right-hand side is zero. This leads to the simple relation:
\begin{eqnarray}\label{intrel}
 \int_{0}^{\infty} dy \;f(y)\ = \frac{\pi }{2}
\end{eqnarray}
This relation is well satisfied numerically for all the mass ratio $r$ that we have investigated.

Another simple relation is obtained by merely setting $x=0$ in Eq.(\ref{eqf}). This gives:
\begin{equation}\label{eqf0}
\frac{R}{2}\;f(0) =1-\frac{2}{\pi} \int_{0}^{\infty} dy \;\frac{f(y)} {1+y^2}
\end{equation}

\section{Very heavy mass}\label{vhma}

In the case where the down-spin particle has a very heavy mass, one has a situation quite close to the case
where this mass is infinite $m\ds = \infty$. In this last case, the physical situation is simpler than in the general
case. Indeed the down particle acts merely as a fixed diffusion center for the two up-spin particles. Numerical calculations
give $a_3=a$ with excellent precision in this case. 

Indeed analytically, since we have then $R=1$ and $R' \rightarrow 0$, 
Eq.(\ref{eqf}) reduces in this case to:
\begin{eqnarray}\label{}
\;\frac{f(x)}{1+\sqrt{1+ x^2}}
    =\frac{1}{1+x^2}-
\frac{2}{\pi} \int_{0}^{\infty} dy \;f(y)
    \,\frac{1}{1+x^2+y^2}
\end{eqnarray}
and we have found that it has the simple, but non-trivial, solution:
\begin{eqnarray}\label{}
f(x)=\frac{1}{1 + x^2}
\end{eqnarray}
since:
\begin{eqnarray}\label{}
 \int_{0}^{\infty} dy\; \frac{1}{1 + y^2}
\,\frac{1}{1+x^2+y^2} = \frac{\pi }{2\, x^2} \left(1-\frac{1}{\sqrt{1+x^2}}\right)= \frac{\pi }{2}\;\frac{1}{1+\sqrt{1+x^2}}
\frac{1}{\sqrt{1+x^2}}
    \end{eqnarray}
Hence we find $a_3=f(0)=1$, which shows that the above conclusion from the numerical results is actually exact.
We can check on this simple solution that Eq.(\ref{intrel}) (and naturally also Eq.(\ref{eqf0})) is indeed satisfied.

This simple but non trivial result can be easily understood if we go back to the ${\bf r}$ space formulation and consider
the wave function of the system. In the limit of a very heavy mass $m_{\downarrow}=\infty$, the spin down particle 
does not move and it acts as a simple diffusion center, located at the origin for convenience, for the two
spin up particles. Since they do not interact, the eigenstates are properly antisymmetrized products of single spin up particle
wave functions. When one spin up particle is far away from the origin, we want to describe a physical situation where one spin
down particle is in the bound state linked to the diffusion center, described by the (non normalized) wave function $\exp(-r_1/a)/r_1$.
The other spin up is in a scattering state corresponding to an incident particle described by a plane wave $\exp (i{\bf k}_0.{\bf r}_2)$
and a scattered particle in a diverging wave $\exp (i k_0. r_2)/r_2$. We are only interested by the s-wave component of this
wave function, which is $\sin(k_0 r_2+\delta _0(k_0))/(k_0 r_2)$, with $\delta _0(k_0)$ the s-wave phase shift.
This is the boundary condition for our scattering problem, but since the spin up particles do not interact this wave function is an
exact eigenstate for any value of ${\bf r}_2$. The full antisymmetric wave function is then:
\begin{eqnarray}
\Psi({\bf r}_1,{\bf r}_2)&=&\frac{\exp(-r_1/a)}{r_1}\,\frac{\sin(k_0 r_2+\delta _0(k_0))}{k_0 r_2}
- \frac{\exp(-r_2/a)}{r_2}\,\frac{\sin(k_0 r_1+\delta _0(k_0))}{k_0 r_1}
\end{eqnarray}
It is already clear at this stage that the scattering amplitude is the same as the one of a single spin up particle scattered by
the diffusion center. The spin up particle in the bound state does not play any role since the two particles do not interact.

If we consider the $k_0 \to 0$ limit, which is the one of interest for us for the scattering length, taking $\delta _0(k_0) \simeq -k_0a$
into account, we have merely
\begin{eqnarray}
\Psi({\bf r}_1,{\bf r}_2)&=&\frac{\exp(-r_1/a)}{r_1}\left(1-\frac{a}{r_2}
\right)-\frac{\exp(-r_2/a)}{r_2}\left(1-\frac{a}{r_1}
\right)\label{eqPsiX}
\end{eqnarray}
Following Skorniakov and Ter-Matirossian \cite{stm}, we define a scattering wave function
$\bar{\chi}(r)={\rm lim}_{r_1 \rightarrow 0}\left(r_1 \Psi({\bf r}_1,{\bf r})\right)$.
From Eq.(\ref{eqPsiX}), we find
\begin{eqnarray}
\bar{\chi}(r)&=&1-\frac{a}{r}+\frac{a}{r}\exp(-r/a)\label{eqfrX}
\end{eqnarray}
The large distance behavior of $\bar{\chi}(r)$ determines the atom-dimer scattering length $a_3$ according to
\begin{eqnarray}
\bar{\chi}(r)&\approx&1-\frac{a_3}{r}
\end{eqnarray}
valid for $r\gg a$. Looking at Eq.(\ref{eqfrX}) in the limit $r\gg a$, we find indeed $a_3=a$.

Let us define $\chi({\bf p})$ as the Fourier transform of $\bar{\chi}(r)$. We know \cite{stm} that $a_3({\bf p})$ and 
$\chi({\bf p})$  are related to each other according to
\begin{eqnarray}
\chi({\bf p})&=&(2\pi)^3\,\delta({\bf p})-4\pi\frac{a_3({\bf p})}{{\bf p}^2}\label{eqfa3}
\end{eqnarray}
If we Fourier transform Eq.(\ref{eqfrX}), we find $\chi({\bf p})=(2\pi)^3\,\delta({\bf p})-4\pi
\frac{1}{a(a^{-2}+{\bf p}^2){\bf p}^2}$. This corresponds to 
\begin{eqnarray}
a_3({\bf p})&=&\frac{1}{a(a^{-2}+{\bf p}^2)}
\end{eqnarray}
which is precisely what we found (translated in non reduced units) as a solution of our integral equation.

The normalisation condition Eq.(\ref{intrel}) can also easily be understood in this limit. Using Eq.(\ref{eqfa3}), we easily find
\begin{eqnarray}
\int_{0}^{\infty}dp\,a_3(p)&=&\frac{\pi}{2}-\frac{\pi}{2}\int\frac{d^3{\bf p}}{(2\pi)^3}\chi({\bf p})\nonumber\\
&=&\frac{\pi}{2}-\frac{\pi}{2}\,\bar{\chi}(r=0)
\end{eqnarray}
From Eq.(\ref{eqfrX}) we can directly check that $\bar{\chi}(r=0)=0$, and therefore $\int_{0}^{\infty}dp\,a_3(p)=\pi/2$. 
However this result is obvious physically and therefore quite general. It states that, when the two up spin particles are
both at the origin (${\bf r}_1={\bf r}_2={\bf 0}$), the wave function $\Psi({\bf r}_1,{\bf r}_2)$ is zero due to Pauli exclusion principle.
Hence it is quite likely that the same explanation for Eq.(\ref{intrel}) holds in the general case where $m\ds$ is not infinite,
and we show explicitly in Appendix \ref{app} that this is indeed the case.

\section{Very light mass}\label{vlm}

This situation turns out to be much more complicated than the preceding one, and it is the essential focus of our paper.

\subsection{Case $m\ds=0$}\label{mzero}

Let us first investigate the limit $m\ds \rightarrow 0$. We will find a fairly singular result, but this will allow us to understand how
to handle the case of a very light mass. For $m\ds \rightarrow 0$, we have $R \rightarrow 0$ and $R' \rightarrow 2$,
so that Eq.(\ref{eqf}) becomes:
\begin{equation}\label{eqfm0}
\int_{0}^{\infty} dy \;\frac{f(y)}{y}\,\ln \frac{1+(x+y)^2}{1+(x-y)^2}=2\pi\,\frac{x}{1+x^2}
\end{equation}

Quite generally, from Eq.(\ref{eqf}), $f(x)$ is an even function of $x$. This allows us to rewrite Eq.(\ref{eqfm0}) as:
\begin{equation}\label{eqfm0a}
-\int_{-\infty}^{\infty} dy \;\frac{f(y)}{y}\,\ln \left[1+(x-y)^2\right]=2\pi\,\frac{x}{1+x^2}
\end{equation}
This convolution product can more easily be handled by integrating by parts. Introducing:
\begin{eqnarray}\label{}
{\cal F}(y)= \int_{-\infty}^{y}dy \;\frac{f(y)}{y}
\end{eqnarray}
which satisfies ${\cal F}(+\infty)=0$ by parity, and is even with respect to $y$, by parts integration gives:
\begin{eqnarray}\label{}
\int_{-\infty}^{\infty} \,dy\,{\cal F}(x-y)\,\frac{y}{1+y^2}=-\,\pi\,\frac{x}{1+x^2}
\end{eqnarray}
This is easily solved by Fourier transform. The Fourier transfom of $x/(1+x^2)$ is:
\begin{eqnarray}\label{}
t(k) \equiv -i\pi \,\mathrm {sign}(k)\;\exp(-|k|)
\end{eqnarray}
but since it appears on both sides of the equation, the result for the Fourier transform $\bar{{\cal F}}(k)$ (which is real and even):
\begin{eqnarray}\label{}
\bar{{\cal F}}(k)= \int_{-\infty}^{\infty} dy\, {\cal F}(y)\, \exp(-iky)
\end{eqnarray}
is immediately seen to be:
\begin{eqnarray}\label{}
\bar{{\cal F}}(k)=-\,\pi 
\end{eqnarray}
Going back to our original function this translates into:
\begin{eqnarray}\label{fsol}
f(x)=-\,\pi \,x\,\delta'(x)
\end{eqnarray}
where the derivative $\delta'$ of the Dirac $\delta$ function is defined as usual by
$ \int_{-\infty}^{\infty}dx\,\delta'(x)\,F(x)=-F'(0)$. Indeed inserting the above result for $f(x)$ into Eq.(\ref{eqfm0a}), it is
immediately seen that it is solution. We can check that relation Eq.(\ref{intrel}) is satisfied since:
\begin{eqnarray}\label{intrelver}
 \int_{0}^{\infty} dy \;f(y)= \frac{1}{2}\;\int_{-\infty}^{\infty} dy \;f(y)=-\frac{\pi }{2} \int_{-\infty}^{\infty}
 \,x\,\delta'(x)=\frac{\pi }{2}
\end{eqnarray}

Actually this solution is not unique since, from Eq.(\ref{eqfm0}), one may add any function extremely localized around the origin
(i.e. a $\delta$-like function), and having a zero total weight. An example of such a function is $\delta(x)+x\,\delta'(x)$.
As a result $\pi \delta(x)$ is also solution, as it is easily directly checked in Eq.(\ref{eqfm0}).
This singular degeneracy will disappear when we consider in the next subsection a very small, but non zero, $m\ds$
and we look for a regular solution. It makes also pointless to discuss the value of the scattering length, although the
fact that $\pi \delta(x)$ is solution makes reasonable the fact that it is infinite in the present limit.

If we take Eq.(\ref{eqf0}) we see that the right-hand side is zero. This looks to be coherent with the fact that $R=0$
in the case we consider. But since we suspect that $f(0)$ goes to infinity, this shows more precisely that the
scattering length does not diverge as fast as $1/R \sim m\us/(2 m\ds)$. This will be confirmed in the next subsection.

Finally we note that the solution Eq.(\ref{fsol}) we have found is completely concentrated at $x=0$. Accordingly we
expect that, for the case of very small $m\ds$ which we investigate in the next subsection, the solution will be
strongly concentrated around $x=0$ and decay very rapidly for large $x$. We remark as a consequence that 
the condition we required for Eq.(\ref{intrel}) to be valid is satisfied.

\subsection{Very small $m\ds$}

Just as in the preceding case we will use a Fourier transform, so it is useful to translate the various quantities and
relations we are interested in. First of all we have for the scattering length:
\begin{eqnarray}\label{}
\bar{a}_3 \equiv \frac{a_3}{a}=f(0)={\rm lim}_{y \rightarrow 0} f(y)
\end{eqnarray}
with
\begin{eqnarray}\label{ft}
f(y)=y\, \frac{d}{dy}{\cal F}(y)
=\int_{-\infty}^{\infty}\frac{dk}{2\pi }\,k\,\bar{{\cal F}}(k) \frac{d}{dk}\exp(iky)=-\int_{-\infty}^{\infty}\frac{dk}{2\pi }\frac{d}{dk}\left(\,k\,\bar{{\cal F}}(k)\right) \exp(iky)
\end{eqnarray}
(to avoid divergences, $k$-integrals must be regularized by putting a factor $\exp(-\eta |k|)$ and then letting $\eta \rightarrow  0_+$) so that:
\begin{eqnarray}\label{condlim}
\bar{a}_3=-\frac{1}{\pi }{\rm lim}_{k \rightarrow +\infty} k\,\bar{{\cal F}}(k)
\end{eqnarray}
since $\bar{{\cal F}}(k)$ is an even function of $k$. In the case of the preceding subsection with $m\ds=0$, this gives indeed $\bar{a}_3=\infty$
because we had $\bar{{\cal F}}(k)=-\,\pi $.

Similarly the relation Eq.(\ref{intrel}) becomes:
\begin{eqnarray}\label{}
\frac{\pi }{2}= \int_{0}^{\infty} dy \;f(y)\ =  \int_{0}^{\infty} dy\,y\, \frac{d}{dy}{\cal F}(y)=- \int_{0}^{\infty} dy\,{\cal F}(y)
\end{eqnarray}
which gives:
\begin{eqnarray}\label{intrelk}
\bar{{\cal F}}(0)=-\pi 
\end{eqnarray}

Let us now expand Eq.(\ref{eqf}) to lowest order in $r=m\ds/m\us$, with $R \simeq 2r$ and $R' \simeq 2(1-r)$. The left hand-side
becomes $R\,f(x)/2$. One may wonder if neglecting entirely the term $Rx^2$ is a proper approximation, since even when $R$
is small, it may become very large if $x$ goes to infinity. However we have stressed that, for small $m\ds$, $f(x)$ is very strongly
peaked around $x=0$ and becomes very small for large $x$. Hence what happens for large $x$ is irrelevant and our approximation
is legitimate. In the right-hand side we expand the logarithm to first order in $R'-2=-2r$. Just as in subsection \ref{mzero} we take
advantage of the even parity of $f(x)$ to extend the integral from $-\infty$ to $\infty$. After multiplication by $\pi R'x$, we obtain:
\begin{eqnarray}\label{eqexpand}
2\pi r \,x\,f(x)=2\pi \,(1-r)\,\frac{x}{1+x^2}+\int_{-\infty}^{\infty} dy \;\frac{f(y)}{y}\,\ln \left[1+(x-y)^2\right]
+2r\,x\,\int_{-\infty}^{\infty} dy \;f(y)\,\frac{1}{1+(x-y)^2}
\end{eqnarray}

Again in the last term the convolution product will give a simple product in Fourier transform. The Fourier transform of $1/(1+x^2)$
is merely $i \,\mathrm {sign}(k)\,t(k)$. Taking into account that the Fourier transform of $f(x)$ is $- d(k\,\bar{{\cal F}}(k))/dk$ from Eq.(\ref{ft}),
we obtain:
\begin{eqnarray}\label{}
t(k)\bar{{\cal F}}(k)+ \pi \,(1-r)\,t(k)+i\pi r\frac{d^2}{dk^2}\left(k\bar{{\cal F}}(k)\right)+
r\frac{d}{dk}\left[t(k)\frac{d}{dk}\left(k\bar{{\cal F}}(k)\right)\right]=0
\end{eqnarray}
where we have restricted ourselves to the case $k>0$, which is enough since $\bar{{\cal F}}(k)$ is even in $k$.
We see immediately that, if we assumed $\bar{{\cal F}}(k)$ to be a constant, the result would still be $\bar{{\cal F}}(k)=-\pi $,
unchanged with respect to the case $m\ds=0$. Introducing accordingly $G(k)=\bar{{\cal F}}(k)+\pi $, performing the
derivations and making use of the explicit expression of $t(k)$, we find the homogeneous differential equation:
\begin{eqnarray}\label{eqG}
r \left[k(1-e^k) \,G''+\left(2(1-e^k)-k\right) G'\right] +G=0
\end{eqnarray}

We see that condition Eq.(\ref{intrelk}), which becomes $G(0)=0$, is automatically satisfied by Eq.(\ref{eqG}).
More precisely, in the vicinity of $k=0$, the two independent solutions of Eq.(\ref{eqG}) behave as $k^{\alpha }$,
with $\alpha =-1 \pm \sqrt{1+1/r}$. Only the solution with $\alpha =-1+ \sqrt{1+1/r}$ satisfies Eq.(\ref{intrelk}),
which provides us with a boundary condition.

Since $r$ is small, this equation is similar to a Schr{\"o}dinger equation for which the WKB approximation could be applied.
It is convenient to make a change of function to bring it precisely under the form of a Schr{\"o}dinger equation. This is done
by setting:
\begin{eqnarray}\label{eqchg}
G(k)=\frac{1}{k\,\sqrt{1-e^{-k}}}\;g(k)
\end{eqnarray}
which leads to:
\begin{eqnarray}\label{eqg}
k(1-e^k) \,g''+\left[1-\frac{k}{2}-\frac{1}{4}\frac{k}{e^k -1}+\frac{1}{r}\right]g=0
\end{eqnarray}
Just as $G(k)$, $g(k)$ must satisfy:
\begin{eqnarray}\label{}
g(0)=0
\end{eqnarray}
On the other hand, because of the term $e^k$ in the coefficient of $g''$, as soon as $k$ is of order of a few units, this coefficient will be large compared to the
coefficient of $g$. More specifically, when $k \gtrsim \ln(1/r)$, Eq.(\ref{eqg}) will reduce to $g''=0$ (we might worry that, in the coefficient of $g$, $-k/2$ will
dominate $1/r$ for large $k$; however when this occurs we will be anyway fully in the regime where $g''=0$ is valid, so this does not change our conclusion).
In this regime the general solution is naturally $g(k)=Ak+B$. However from Eq.(\ref{condlim}) we know that, for large $k$, $\bar{{\cal F}}(k) \simeq -\pi a_3/k$.
This implies $G(k) \simeq \pi -\pi a_3/k$ and from Eq.(\ref{eqchg}) $g(k) \simeq \pi k -\pi a_3$. Hence we know that, for large $k$, $g(k)$ has an asymptote
with slope $A=\pi $. This fixes the coefficient of the solution of the homogeneous equation Eq.(\ref{eqg}) satisfying $g(0)=0$. Hence we have a unique
solution as it should be, and the constant $B=-\pi a_3$ will gives us the scattering length. Equivalently we can say that, if we take a solution satisfying
$g(0)=0$, its asymptotic behaviour for large $k$ is $g(k) \simeq Ak+B$ and we have $a_3=-B/A$.

Now, with the above discussion, we can further simplify Eq.(\ref{eqg}) taking into account that we are only interested in this equation when $r$ is small.
In this case, when $k$ is not very large, the coefficient of $g$ is essentially equal to $1/r$. On the other hand when $k$ is large enough so that this is
no longer valid, the equation reduces to $g''=0$. Hence in all cases Eq.(\ref{eqg}) may be simplified into:
\begin{eqnarray}\label{eqwkb}
\epsilon ^2 g''(k)=\frac{1}{k(e^k-1)}\,g(k) \equiv Q(k) g(k)\hspace{15mm} Q(k)=\frac{1}{k(e^{k}-1)}
\end{eqnarray}
where we have set $r=\epsilon ^2$.

\subsection{WKB solution}
Since we are in a situation where $\epsilon ^2$ is very small, we can solve Eq.(\ref{eqwkb}) by the WKB method. The solution which satisfies $g(0)=0$
is, within a multiplicative constant:
\begin{eqnarray}\label{solwkb}
g(k)=\frac{1}{[Q(k)]^{1/4}}\; \exp\left[-\frac{1}{\epsilon } \int_{k}^{\infty}dx\,\sqrt{Q(x)}\right]
\end{eqnarray}
where the arbitrary choice of the upper bound of the integral modifies only the arbitrary multiplicative constant.
We note that, for $k \to 0$, $Q(k) \simeq 1/k^2$, and the solution Eq.(\ref{solwkb}) behaves as $g(k) \sim k^{\frac{1}{2}+\frac{1}{\epsilon }}$.
This is naturally in agreement with what can be obtained directly from Eq.(\ref{eqwkb}), and this corresponds to what has been found for
Eq.(\ref{eqG}) for small $\epsilon $, taking the change Eq.(\ref{eqchg}) into account.

Unfortunately we can not extract from Eq.(\ref{solwkb}) the asymptotic behaviour of $g(k)$ for large $k$ because in this limit $Q(k) \rightarrow 0$
and we have to deal with a turning point for our Eq.(\ref{eqwkb}) which is located at $k=\infty$. Nevertheless for large $k$ we have $Q(k) \simeq
e^{-k}/k$ and Eq.(\ref{eqwkb}) becomes:
\begin{eqnarray}\label{eqwkbgdk}
\epsilon ^2 g''=\frac{e^{-k}}{k}\,g
\end{eqnarray}
and its solution Eq.(\ref{solwkb}) simplifies somewhat into:
\begin{eqnarray}\label{solgdk}
g(k)=k^{1/4}\,e^{k/4}\; \exp\left[-\frac{1}{\epsilon } \int_{k}^{\infty}dx\,\frac{e^{-x/2}}{\sqrt{x}}\right]
\end{eqnarray}
As for the standard treatment of the Schr\"{o}dinger equation within the WKB approximation, we have to solve directly Eq.(\ref{eqwkbgdk}) near
the turning point and match the solution to Eq.(\ref{solgdk}). Since the turning point is at infinity it is more convenient to bring it to zero by the
additional change of variable:
\begin{eqnarray}\label{}
z=2\,e^{-\frac{k}{2}}
\end{eqnarray}
which transforms Eq.(\ref{eqwkbgdk}) into an equation for $\bar{g}(z)=g(k)$, namely:
\begin{eqnarray}\label{eqturn}
z\,\frac{d^2\bar{g}(z)}{dz^2}+\frac{d\bar{g}(z)}{dz}-\frac{z}{L(z)}\, \bar{g}(z)=0
\end{eqnarray}
with $L(z)=2\epsilon ^2 \ln(2/z)$ and the solution Eq.(\ref{solgdk}) reads now within an arbitrary multiplicative constant:
\begin{eqnarray}\label{solz}
\bar{g}(z)=\frac{[L(z)]^{1/4}}{\sqrt{z}} \;\exp\left[- \int_{0}^{z}dx\,\frac{1}{[L(x)]^{1/2}}\right]
\end{eqnarray}
In contrast with the standard treatment of the turning point in the Schr\"{o}dinger equation,
where the solutions are the Airy functions with known mathematical properties,
we do not know the solutions of Eq.(\ref{eqturn}) and we would have to find their mathematical properties.
However we may notice (and this could be a first step in a mathematical study) that, for small $z$, $L(z)$
is a quite slowly varying function of $z$, compared to $z$ or $z^2$. In a first step, we may treat it as a constant
$L(z) \simeq L$. Then the further change of variable $z=\sqrt{L}\,x$ brings Eq.(\ref{eqturn}) into:
\begin{eqnarray}\label{eqturnx}
x\,\frac{d^2\tilde{g}(x)}{dx^2}+\frac{d\tilde{g}(x)}{dx}-x \tilde{g}(x)=0
\end{eqnarray}
for $\tilde{g}(x)=\bar{g}(z)$. The solutions of this equation are \cite{gr} the well known Bessel functions $I_0(x)$ and $K_0(x)$.
The form of the solution Eq.(\ref{solz}) for large $z$, which becomes now within an arbitrary multiplicative constant:
\begin{eqnarray}\label{solx}
\tilde{g}(x)=\frac{1}{\sqrt{x}}\,e^{-x}
\end{eqnarray}
serves as a boundary condition to find the proper solution of Eq.(\ref{eqturnx}). Since for large $x$, we have $I_0(x)
\simeq e^x/\sqrt{2\pi x}$ while $K_0(x) \simeq \sqrt{\pi/2 x}\,e^{-x}$, we see from Eq.(\ref{solx}) that the solution we are looking for is just
$\tilde{g}(x)=K_0(x)$. On the other hand it is known \cite{gr} that, for small $x$, $K_0(x) \simeq -\ln x+\ln 2 - C$ where $C=0.577215...$
is the Euler constant. Going back to the variable $k$ through $x=2\,e^{-k/2}/\epsilon \sqrt{k}$, we obtain:
\begin{eqnarray}\label{asympg}
g(k) \simeq \frac{k}{2}+\ln \epsilon -C+\frac{1}{2}\,\ln k
\end{eqnarray}
The last term, proportional to $\ln k$, is not consistent with Eq.(\ref{eqwkbgdk}) and its presence is due to our above approximation
of treating the logarithm $L(z)$ as essentially constant. However if we stay coherent with this approximation and notice that the domain
we consider for the behaviour of $K_0(x)$ is $x\sim1$, we have to dominant order in this domain $k\simeq \ln (1/\epsilon ^2)$ and
we have to replace coherently $\ln k$ by $\ln (\ln (1/\epsilon ^2)$. As we have seen ${\bar a}_3$ is obtained from the ratio of the constant
to the coefficient of $k$ in Eq.(\ref{asympg}). From $\epsilon =\sqrt{m\ds /m\us}$ this leads to:
\begin{eqnarray}\label{eqa3}
{\bar a}_3 \simeq \ln (m\us /m\ds)-\ln (\ln (m\us /m\ds))+2C
\end{eqnarray}
Hence to dominant order we find ${\bar a}_3 \simeq \ln (m\us /m\ds)$, which is in agreement with the estimate by Petrov \cite{petr} in the large $m\us /m\ds$ regime.
However our handling of the $\ln k$ term as a quasi constant may look somewhat uncertain and one may be more uncertain about the last
two terms of our result Eq.(\ref{eqa3}). Therefore we present in the next section
an alternative derivation of the scattering length from Eq.(\ref{eqwkb}) which will give us more confidence in our result Eq.(\ref{eqa3})
and will even allow us to go beyond this result.

\subsection{Alternative derivation of the scattering length}

In the preceding section we have obtained our result by matching the WKB solution at infinity, which requires a somewhat uneasy handling. 
Here, in order to obtain the scattering length, we use another method which avoids this problem, or rather puts it on a much more secure basis. 
It makes use of the fact that $Q(k)$ goes very rapidly to zero when $k$ goes to infinity. Hence we can integrate Eq.(\ref{eqwkb}), making use of the boundary
conditions $g(0)=0$ and $g'(0)=0$ which results directly from the $g(k) \sim k^{\frac{1}{2}+\frac{1}{\epsilon }}$ behaviour we
have found in the preceding section. A first integration gives:
\begin{eqnarray}\label{eqwkb1}
\epsilon ^2 g'(k)= \int_{0}^{k}dx\,Q(x)\,g(x)
\end{eqnarray}
Another integration, followed by a by parts integration leads to:
\begin{eqnarray}\label{eqwkb2}
\epsilon ^2 g(k)=k \int_{0}^{k}dx\,Q(x)\,g(x)- \int_{0}^{k}dx\,x\,Q(x)\,g(x)
\end{eqnarray}
Since $Q(k)$ goes very rapidly to zero when $k$ goes to infinity, whereas we know that $g(k)=Ak+B$ for large $k$, we may replace the
upper bounds of the integrals by $+\infty$ as soon as $k$ is large. This gives in this large $k$ regime:
\begin{eqnarray}\label{eqwkb3}
\epsilon ^2 g(k)=k \int_{0}^{\infty}dx\,Q(x)\,g(x)- \int_{0}^{\infty}dx\,x\,Q(x)\,g(x) \equiv {\bar A}k- {\bar B}
\end{eqnarray}
which displays precisely the expected asymptotic behaviour. The scattering length is then obtained from $a_3={\bar B}/{\bar A}$.
At first this seems useless since the integrals imply the knowledge $g(k)$. However, since $Q(k)$ goes very rapidly to zero, we can use
one of the approximate solutions which have been discussed above and which would be valid for the whole range of integration. 
A first idea is to use the WKB solution Eq.(\ref{solwkb}) which leads, within an unimportant multiplicative constant, to:
\begin{eqnarray}\label{eqA}
{\bar A}&=&\int_{0}^{\infty}dk\,\left[Q(k)\right]^{3/4}\,\exp\left[-\frac{1}{\epsilon } \int_{k}^{\infty}dx\,\sqrt{Q(x)}\right] \\
{\bar B}&=&\int_{0}^{\infty}dk\,k\,\left[Q(k)\right]^{3/4}\,\exp\left[-\frac{1}{\epsilon } \int_{k}^{\infty}dx\,\sqrt{Q(x)}\right]
\label{eqB}
\end{eqnarray}
Since the WKB solution increases rapidly with $k$ while $Q(k)$ decreases rapidly, a further idea is to use a saddle point estimate.
For ${\bar B}$ for example the saddle point $k_0$ corresponds to a maximum of the integrant of Eq.(\ref{eqB}), or rather its logarithm.
This gives for $k_0$ the equation:
\begin{eqnarray}\label{}
\frac{1}{\epsilon }=\frac{3}{4\sqrt{k_0}}\,e^{\frac{k_0}{2}}
\end{eqnarray}
where we have already used the fact that $k_0$ is large. To dominant order this gives $k_0 \simeq 2 \ln (1/\epsilon )=\ln (m\us /m\ds)$.
To dominant order the saddle point for ${\bar A}$ is located at the same place, because ${\bar A}$ and ${\bar B}$ differ only by the factor
$k$ in the integrants. For the same reason we obtain to dominant order $a_3={\bar B}/{\bar A} \simeq k_0 \simeq \ln (m\us /m\ds)$, which
allows already to recover the dominant order of our result Eq.(\ref{eqa3}). Unfortunately this result for $k_0$ shows that the saddle point
is located at a $k$ value where the argument of the exponential in the WKB solution Eq.(\ref{solwkb}) is of order one. This implies that
the WKB solution is no longer reliable because this $k_0$ is too large and we are already not far enough from the turning point located
at infinity.

Fortunately we have found an approximate solution of Eq.(\ref{eqwkb}) which matches the WKB solution to the solution at infinity,
namely $K_0(x)$ with $x=2 e^{-k/2}/\epsilon \sqrt{k}$. This solution is only approximate because we have treated $k$ as a quasi constant,
compared to $e^{-k}$. However the fact that the integrants for ${\bar A}$ and ${\bar B}$ are sharply peaked, allowing as we have seen the use of a saddle point
evaluation, makes this treatment of $k$ as a quasi constant a very good approximation for our purpose. In particular the sharp decrease
of $Q(k)$ makes that we do not have to worry about very large values of $k$.

Performing the change of variable $x=2 e^{-k/2}/\epsilon \sqrt{k}$ in ${\bar A}$ and ${\bar B}$, as given in Eq.(\ref{eqwkb3}), with $k$ quasi constant,
we find:
\begin{eqnarray}\label{eqAa}
{\bar A}&=&\frac{\epsilon ^2}{2}\int_{0}^{\infty}dx\,x\,K_0(x) \\
{\bar B}&=&\frac{\epsilon ^2}{2}\int_{0}^{\infty}dx\,x\,k(x)\,K_0(x)
\label{eqBb}
\end{eqnarray}
where $k(x)$ is the inverse function of $x=2 e^{-k/2}/\epsilon \sqrt{k}$. It satisfies:
\begin{eqnarray}\label{invf}
k(x)=\ln \frac{1}{\epsilon ^2}-2 \ln x + 2 \ln 2 - \ln k(x)
\end{eqnarray}
Actually we do not need a saddle point calculation since we have the exact results \cite{gr}:
\begin{eqnarray}\label{}
\int_{0}^{\infty}&dx&\,x\,K_0(x)=1 \\
\int_{0}^{\infty}&dx&\,x\,\ln x\,K_0(x)=\ln 2 - C
\end{eqnarray}
Since in the integral we have $x \sim 1$ because $K_0(x)$ decreases exponentially for large $x$, we have to dominant order $k(x) \simeq \ln(1/\epsilon ^2)$
and we may replace $k(x)$ by this approximate value in the last term in the right-hand side of Eq.(\ref{invf}). This leads us exactly again to 
Eq.(\ref{eqa3}) for the expression of ${\bar a}_3$.

\begin{figure}
\centering
{\includegraphics[width=110mm]{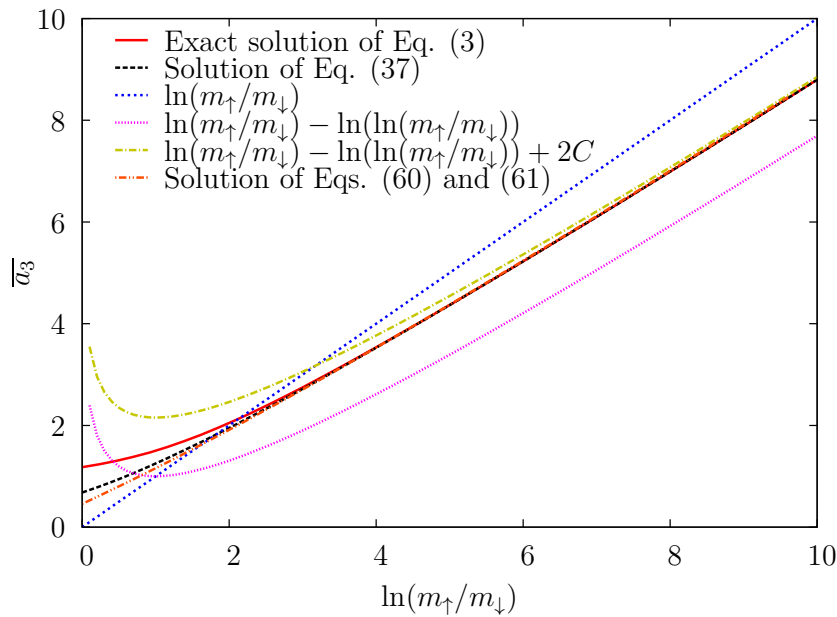}}
\caption{(color online). Reduced fermion-dimer scattering length $\bar{a}_3=a_3/a$ as a function of the mass ratio $m\us /m\ds$.
Full line: exact solution from Eq.(\ref{eqf}). Dashed line: exact solution of Eq.(\ref{eqwkb}) corresponding to the first order approximation of Eq.(\ref{eqf})
for small $m\ds /m\us$. Dotted line: dominant result $\ln (m\us /m\ds)$. Small dotted line: inclusion of the subdominant contribution -$\ln(\ln (m\us /m\ds))$.
Dashed-dotted line: full asymptotic result Eq.(\ref{eqa3}). Dashed-double-dotted line: parametric representation Eq.(\ref{eqs2}) and Eq.(\ref{eqs3}).}
\label{fig1}
\end{figure}

We may even try to improve on this result by refining our handling of the last term in Eq.(\ref{invf}). We could just iterate our procedure.
However this would produce new integrals to evaluate and there is no justification to go to such a refinement. On the other hand, since we have seen that
$x \sim 1$ for the dominant range of integration, we may just set $\ln x \simeq 0$. This leads to replace, in the last term in the right-hand 
side of Eq.(\ref{invf}), $k(x)$ by ${\bar k}_0$. This gives for ${\bar k}_0$ the equation:
\begin{eqnarray}\label{eqs1}
\ln (m\us /m\ds)={\bar k}_0+\ln {\bar k}_0- 2 \ln 2
\end{eqnarray}
while ${\bar a}_3$ is still obtained from Eq.(\ref{invf}) by replacing $k(x)$ by ${\bar k}_0$ in the last term in the right-hand side, which yields:
\begin{eqnarray}\label{eqs2}
{\bar a}_3 = \ln (m\us /m\ds)- \ln {\bar k}_0+2C
\end{eqnarray}
This gives us a parametric representation of ${\bar a}_3$ in terms of $\ln (m\us /m\ds)$, with ${\bar k}_0$ as parameter.
This is easily plotted. The result has the pleasant feature of eliminating the divergent behaviour of Eq.(\ref{eqa3}) for
small $\ln (m\us /m\ds)$ (which is nevertheless not at all disturbing since we are exploring actually the large
$\ln (m\us /m\ds)$ domain). However the result is slightly less satisfactory for large $\ln (m\us /m\ds)$. This can be
improved by noticing that replacing $\ln x$ by $0$ is just a first reasonable try, but that the actual result is rather
obtained by replacing it by a constant of order $1$. Hence if we replace Eq.(\ref{eqs1}) by:
\begin{eqnarray}\label{eqs3}
\ln (m\us /m\ds)={\bar k}_0+\ln {\bar k}_0- F
\end{eqnarray}
and treat $F$ as a fitting parameter, we may hope to obtain a better result. Indeed 
we obtain an excellent result over the whole range of variation of $\ln (m\us /m\ds)$ for $F=2.7$ as it can be seen in Fig.~\ref{fig1}

\subsection{Comparison with numerical results}\label{}

As we have mentioned at the beginning, it is naturally easy to find  at any stage the numerical answers to the questions we raise.
For example we display in Fig.~\ref{fig1} the numerical solution of the exact problem Eq.(\ref{eqf}) giving ${\bar a}_3$ as a function of $\ln (m\us /m\ds)$. We can also solve numerically for ${\bar a}_3$ the equation Eq.(\ref{eqwkb}) we have obtained for large $m\us /m\ds$. The result is also displayed 
in Fig.~\ref{fig1}. As it could have been expected the difference between the exact solution and this approximate solution of large $m\us /m\ds$ becomes quite small for $\ln (m\us /m\ds) \sim 2$, i.e. $m\us /m\ds \sim 7$. We have then shown the lowest order approximation 
${\bar a}_3 = \ln (m\us /m\ds)$ which is a straight line on this graph and which is clearly not so satisfactory. Next we have plotted the first two terms
of Eq.(\ref{eqa3}). It is clear that the $-\ln (\ln (m\us /m\ds))$ corrective term leads to the proper shape for large $m\us /m\ds$. Finally
plotting the whole Eq.(\ref{eqa3}) shows that the constant $2C$ leads to a perfect agreement with the exact result for large $m\us /m\ds$. 
However for low $m\us /m\ds$ this result departs strongly both from the exact one and from the exact solution of equation Eq.(\ref{eqwkb}), due to the
divergence of the $-\ln (\ln (m\us /m\ds))$ term. This deficiency is eliminated by the parametric solution Eq.(\ref{eqs2}) and Eq.(\ref{eqs3}),
which gives an agreement with the solution of Eq.(\ref{eqwkb}) which almost perfect over the whole range $m\us /m\ds > 1$.

\subsection{The function $\bar{a}_3(\mathbf{\bar{p}})$}\label{}

\begin{figure}
\centering
\includegraphics[width=110mm]{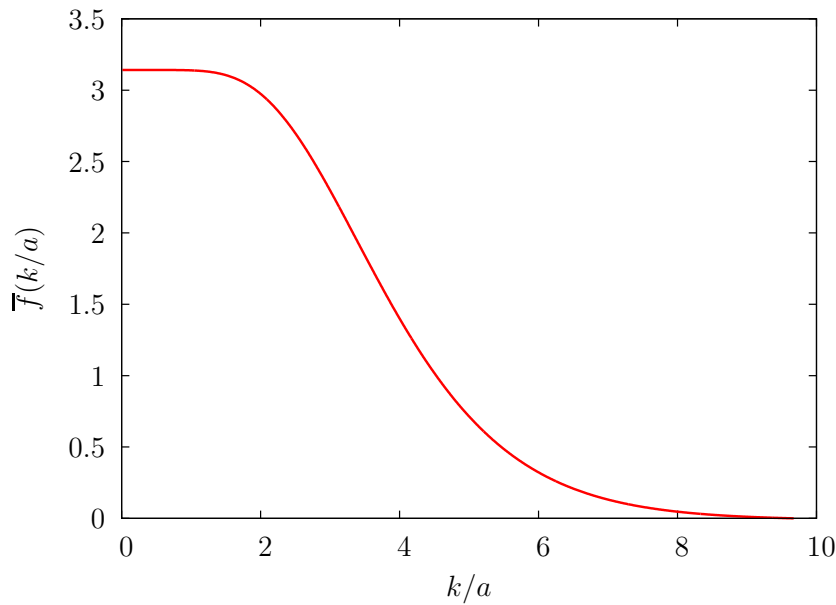}
\caption{(color online). Solution of Eq.(\ref{fbar}) for $m\us /m\ds=10^2$. To be explicit we use non reduced units for $k$, which has the dimension of a length
and is expressed naturally in terms of the scattering length $a$. Note the very flat part for small $k/a$. }
\label{fig2}
\end{figure}

Finally, it is of interest to consider the approximate solution $\bar{a}_3(\mathbf{\bar{p}})$, or $f(x)$ with our simpler notations,
we find for the basic integral equation Eq.(\ref{a_3}). From Eq.(\ref{ft}) its Fourier transform $\bar{f}(k)$ is:
\begin{eqnarray}\label{fbar}
\bar{f}(k)=-\frac{d}{dk}\left(\,k\,\bar{{\cal F}}(k)\right)=\frac{d}{dk}\left(\pi k-\frac{g(k)}{\sqrt{1-e^{-k}}}\right)
\end{eqnarray}
which is easily calculated numerically. An example is given in Fig.~\ref{fig2}. It is peaked around $k=0$ and has a width of order $k_0 \simeq \ln (m\us /m\ds)$
as it can be seen from the WKB expression Eq.(\ref{solwkb}) for $g(k)$. Hence $f(x)$ is peaked around $x=0$ with a typical width $1/\ln (m\us /m\ds)$
and is accordingly very narrow, as expected. One checks that Eq.(\ref{fbar}) gives $f(0)={\bar a}_3$ from the asymptotic behaviour of $g(k)$.
The large $x$ is behaviour is obtained from the small $k$ dependence of $\bar{f}(k)$,
which is $\bar{f}(k) \simeq \pi - k^{-1/2}g(k)$ with $g(k) \sim k^{\frac{1}{2}+\frac{1}{\epsilon }}$ as we have seen. This leads to
an extremely fast decrease $f(x) \sim 1/x^{1+1/\epsilon }$ at large $x$. 

However we are essentially interested in the domain
where $f(x)$ is clearly different from zero, which is obtained for fairly small values of $x$ since $f(x)$ is narrow. These will come from
$\bar{f}(k)$ with fairly large values of $k$. In this range we can set in Eq.(\ref{fbar}) $1-e^{-k} \simeq 1$ and take the approximate solution
$g(k)=2\pi K_0(2e^{-k/2}/(\epsilon \sqrt{k}))$ where the prefactor provides the correct asymptotic behaviour for $g(k)$. On the other hand
if $x \neq 0$ the first term in the right-hand side of Eq.(\ref{fbar}) gives no contribution. Taking into account the even parity of
$\bar{f}(k)$, integrating by parts and taking the complex conjugate we obtain:
\begin{eqnarray}\label{}
f(x)=2x\,{\rm Im}\, \int_{0}^{\infty}dk\,e^{-ikx}\,K_0(\frac{2e^{-k/2}}{\epsilon \sqrt{k}})
\end{eqnarray}
For small $k$ the Bessel function $K_0$ goes very rapidly to zero, while for very large $k$ the Bessel function varies slowly while
the oscillating factor $e^{ikx}$ leads to destructive interferences. Hence the dominant contribution arises when the argument of 
the Bessel function is of order unity, that is again when $k$ is of order $k_0 \simeq \ln (m\us /m\ds)=2 \ln (1/\epsilon) $. Hence we
can again treat in the argument of the Bessel function $k$ as a quasi constant and write $\sqrt{k}\simeq \sqrt{2 \ln (1/\epsilon)}$.
Taking $t=2e^{-k/2}/(\epsilon \sqrt{2 \ln (1/\epsilon)})$ as new variable, and making use of \cite{gr} $ \int_{0}^{\infty}dt\,t^{\mu}\,K_0(t)=
2^{\mu -1}\,\Gamma^2((1+\mu )/2)$, where $\Gamma(x)$ is the Euler function, we obtain:
\begin{eqnarray}\label{fK}
f(x)=x\,{\rm Im}\left\{\left[(m\ds /m\us)\ln (m\us /m\ds)\right]^{ix}\Gamma^2(ix)\right\}
\end{eqnarray}
Since \cite{gr}, for $z \to 0$, we have $\Gamma(z) \simeq 1/z - C$, this result gives back exactly Eq.(\ref{eqa3}) for ${\bar a}_3=f(0)$,
which shows the coherence of our approximate treatment.

\begin{figure}
\centering
\includegraphics[width=110mm]{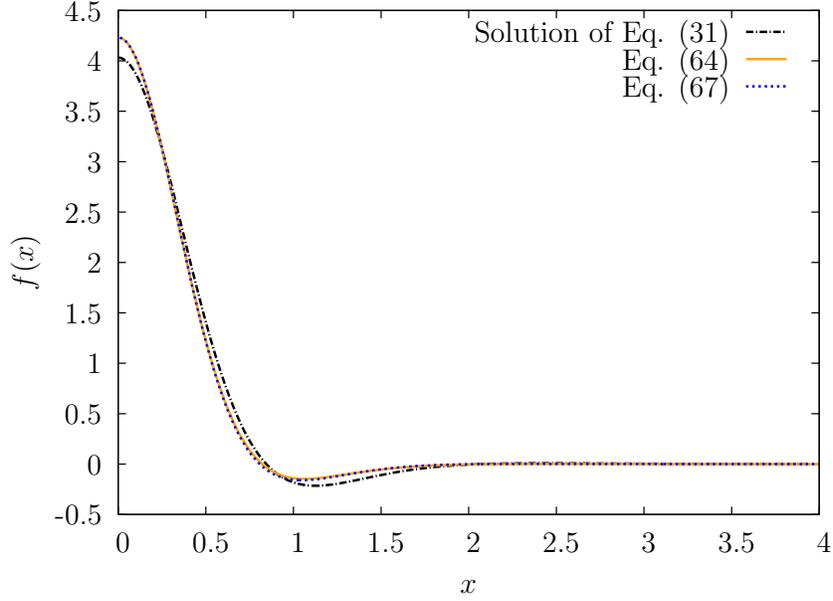}
\caption{(color online). For $m\us /m\ds=10^2$ corresponding to $b\simeq 3.08$,  full line: Eq.(\ref{fK}), dotted line: Eq.(\ref{fKp}),
dashed-dotted line: exact numerical solution of Eq.(\ref{eqexpand}).}
\label{fig3}
\end{figure}

\begin{figure}
\centering
\includegraphics[width=110mm]{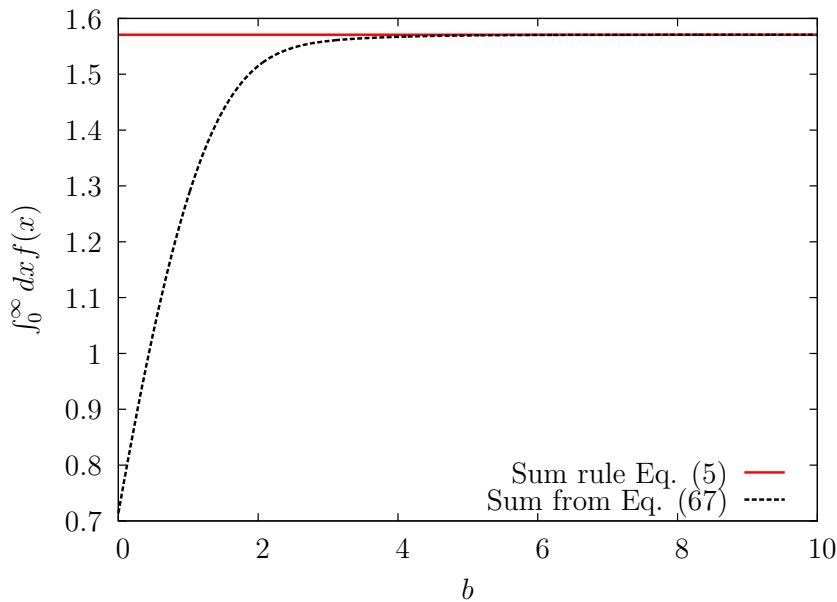}
\caption{(color online). Integral of Eq.(\ref{fKp}) as a function of $b= \ln(m\us /m\ds)-\ln(\ln (m\us /m\ds))$ compared to the exact result $\pi /2$ from
the sum rule Eq.(\ref{intrel}).}
\label{fig4}
\end{figure}

It is possible to write Eq.(\ref{fK}) in a simpler form at the price of a small approximation, by making use of the Stirling expansion of
the Euler function $\ln\Gamma(z) \simeq (z-1/2)\ln z -z+(1/2) \ln(2\pi )$. Writing $\Gamma^2(ix)=|\Gamma(ix)|^2\,(\Gamma(ix)/\Gamma(-ix))$
where \cite{gr} $|\Gamma(ix)|^2=\pi /[x\,\sinh(\pi x)]$, with $\Gamma(ix)/\Gamma(-ix)=-\Gamma(1+ix)/\Gamma(1-ix)$ and making use of the 
Stirling expansion for $\Gamma(1 \pm ix)$, we find:
\begin{eqnarray}\label{ffinal}
f(x) \simeq \frac{\pi }{\sinh(\pi x)}\,\sin\left[\left(b+2-\ln(1+x^2)-\frac{\arctan x}{x}\right)x\right]
\end{eqnarray}
with the notation:
\begin{eqnarray}\label{}
b \equiv \ln(m\us /m\ds)-\ln(\ln (m\us /m\ds))
\end{eqnarray}
This result is an excellent approximation for Eq.(\ref{fK}), except in the vicinity of $x=0$ because it gives
$f(0)=b+1$ instead of $f(0)=b+2C$, which is still a very good approximation since $2C \simeq 1.154$. This small deficiency comes
from the Stirling expansion in the vicinity of $z=1$. We may correct for it by replacing $2$ by $1+2C$. This leads to:
\begin{eqnarray}\label{fKp}
f(x) \simeq \frac{\pi }{\sinh(\pi x)}\,\sin\left[\left(b+1+2C-\ln(1+x^2)-\frac{\arctan x}{x}\right)x\right]
\end{eqnarray}
which is almost indistinguishable from Eq.(\ref{fK}), for any sizeable value of $b$. This can be seen from Fig.~\ref{fig3}
where, for $m\us /m\ds=10^2$ corresponding to $b\simeq 3.08$, we have plotted both Eq.(\ref{fK}) and Eq.(\ref{fKp}).
They are undistinguishable on the figure. We have also plotted the exact solution of the first order approximation 
Eq.(\ref{eqexpand}), for small $m\ds /m\us$, of the exact starting equation Eq.(\ref{eqf}). The agreement is very good.
The only sizeable difference is again in the vicinity of $x=0$ because the asymptotic form Eq.(\ref{eqa3}) ${\bar a}_3=b+2C$
is, for this value of $m\us /m\ds=10^2$, still slightly different from the result obtained from Eq.(\ref{eqexpand}) (or equivalently 
Eq.(\ref{eqwkb})) and Eq.(\ref{eqf}), as it is seen from Fig.~\ref{fig1}. This could be completely corrected by replacing, in Eq.(\ref{fKp}), 
$b+2C$ by the exact ${\bar a}_3$. This would provide an excellent approximation to the exact solution of the original Eq.(\ref{eqf}).
Finally we have also checked that this expression Eq.(\ref{fKp}) satisfies the sum rule Eq.(\ref{intrel}) 
with a remarquable precision as it can be seen from Fig.~\ref{fig4}.

This explicit result Eq.(\ref{fKp}) shows that
$f(x)$ decreases very rapidly, with an oscillatory behaviour in an exponentially decreasing envelope. This function is almost zero after the first zero,
except for a small oscillation just after this zero.
For large $b$, this zero is small and is given essentially by $x_0=\pi /(b+2C)=\pi /{\bar a}_3$. Hence we find explicitely a result coherent
with our starting hypothesis, namely that $f(x)$ gets very narrow for large ${\bar a}_3$. Actually the behaviour resulting from Eq.(\ref{fK}) 
for extremely large $x$ is not the correct one because we have made approximations for the low $k$ behaviour of $\bar{f}(k)$. As we mentioned
above this behaviour is rather $f(x) \sim 1/x^{1+1/\epsilon }$ without any oscillation. 
However since $f(x)$ is anyway extremely small in this regime, this is completely unimportant in practice.
We note that all our above analytical findings are in excellent agreement with the very recent numerical work of Iskin \cite{isk}.
A final point, also in agreement with the work of Iskin \cite{isk} is that the first zero disappears (by coalescence with the second one) 
when $b$ decreases. This feature is fairly natural
since, when $m\ds$ goes to infinity, the solution is the lorentzian found in section \ref{vhma} which has no zero. Indeed the argument
of the sinus in Eq.(\ref{fKp}) is an increasing function of $x$ for small $x$, but decreases for large $x$. The first zero disappears
when this argument reaches its maximum as a function of $x$ and is equal to $\pi$ at this maximum. This occurs for $x_0=2.33$
and $b=1.55$, corresponding to $m\us /m\ds=11.6$. This feature is again in very nice agreement with the full numerical solution \cite{isk} of
Eq.(\ref{eqf}), although it is not very good quantitatively since the zero is found to disappear for $m\us /m\ds \simeq 7$ corresponding to $b=1.28$.
This small quantitative disagreement is expected since this mass ratio does not correspond to the full asymptotic regime of very small
$m\ds$.

We note finally that, if we are not interested in the disappearance of the zeros and rather concentrate on the regime where  ${\bar a}_3$ is quite large,
$f(x)$ can be further simplified into a very simple expression. Indeed we may assume that $x$ is fairly small since otherwise, from Eq.(\ref{fKp}), 
$f(x)$ is extremely small, compared to its $x=0$ value, and its specific expression becomes unimportant. In this case we have $\ln(1+x^2) \simeq 0$ and $\arctan x/x \simeq 1$. Taking
into account  ${\bar a}_3=b+2C$, Eq.(\ref{fKp}) reduces to:
\begin{eqnarray}\label{fKpsimple}
f(x)= \frac{\pi }{\sinh(\pi x)}\,\sin({\bar a}_3 x)
\end{eqnarray}
First this expression satisfies $f(0)={\bar a}_3$. Then with this very simple form the integral 
in the sum rule Eq.(\ref{intrel}) can be calculated analytically. One finds $(\pi /2) \tanh ({\bar a}_3/2)$
which, for large ${\bar a}_3$, is $\pi /2$ as it should be. Furthermore when the right-hand side of Eq.(\ref{eqf0}) is calculated analytically
one finds, to dominant order, the result ${\bar a}_3\,e^{-{\bar a}_3}$. This is in perfect agreement with the left-hand side of Eq.(\ref{eqf0})
since $f(0)={\bar a}_3$ and, to dominant order, $R/2=r=m\ds / m\us \simeq e^{-{\bar a}_3}$ from Eq.(\ref{eqa3}). 
Hence, since ${\bar a}_3$
is large, we have an almost complete cancellation of the two terms in the right-hand side of Eq.(\ref{eqf0}) whereas a simple order of
magnitude evaluation would rather give a comparatively much larger result of order $(1/{\bar a}_3)^2$. It is thus quite remarkable that the non trivial
property Eq.(\ref{eqf0}) is satisfied by the simple expression Eq.(\ref{fKpsimple}).

\section{Conclusion}

In this paper we have considered the problem of the scattering length $a_3$ for a fermion of mass $m\us$ colliding with a dimer, 
formed from a fermion identical to the incident one and another different fermion of mass $m\ds$. 
The only scattering parameter in this problem is the scattering length $a$ between the
different fermions. We have been interested in the way in which this scattering length depends on the mass
ratio $m\ds/m\us$ between the different fermions. The experimental investigation of these kind of systems is under 
current development in several laboratories
in the world. While the answer to this problem is obtained from a generalization of an integral equation first explored by Skorniakov and 
Ter-Martirosian \cite{stm} (STM), which has been fully investigated numerically, we have been interested in an analytical investigation 
of this problem and this equation in order to gain additional insight and open possible roads for more complex problems.

We have been able to perform this investigation in the limiting cases where the mass ratio is very large or very small. When the mass $m\ds$ of the
lonely fermion is very large, the situation is very similar to the one where it is infinite and where the corresponding atom behave as a fixed scattering
center. In this case we have found the analytical solution of the STM integral equation. The corresponding scattering length is merely $a$.
The existence of another fermion in a bound state (i.e. the existence of the dimer) is unimportant and physically everything happens as if the fermion
is merely scattering on the fixed center. It is noteworthy that the equal mass case $m\ds=m\us$, where $a_3/a \simeq 1.18$, is fairly near
this  limiting case.

When the mass $m\ds$ is very small, the atom-dimer scattering length $a_3$ goes to infinity. We have found that in this situation the width of the unknown function in the generalized STM equation becomes very narrow around the origin. We have used this feature to expand properly the 
STM equation in this case. We have in this way obtained the explicit asymptotic behaviour of $a_3$ in this large $m\us/m\ds$ regime. 
We have found:
\begin{eqnarray}\label{eqa3concl}
\frac{{a}_3}{a} \simeq \ln (m\us /m\ds)-\ln (\ln (m\us /m\ds))+2C
\end{eqnarray}
where $C$ is the Euler constant. In addition we have found a parametric representation of $a_3$ in terms of $m\us/m\ds$ 
which is even more accurate at lower $m\us/m\ds$,
and gives an excellent agreement with the exact numerical results over the whole range $m\us/m\ds \gtrsim 7$.
At the same time we have found an excellent approximate analytical solution for the unknown function in the STM equation
in this same regime. In particular this solution displays the same appearance of zeros and oscillations as it is found in the numerical
results of the exact equation for $m\us/m\ds \gtrsim 7$. This is in contrast with the shape of this solution for lower $m\us/m\ds$
which is merely found to decrease monotously to zero.

\appendix

\section{Interpretation of the integral relation Eq.\ref{intrel}}\label{app}
In this Appendix, we show in details the connection between the integral relation Eq.(\ref{intrel})
and the fact that two spin up particles can not be at the same position. In order to do so, we need to make the connection between the function $\bar{a}_3(\bar{p})$ of Eq.(\ref{a_3}) and the wave function $f({\bf r})$ studied by Petrov \cite{petr}. We keep his notations, so we emphasize that in this Appendix, contrary to the rest of the article, $f$ is a wavefunction defined in Eq.(\ref{eqdeff}) below.

We first recall the notations and results of \cite{petr}. The system we study is made of two spin up particles (positions ${\bf x}_1$ and ${\bf x}_2$) and a single spin down particle (position ${\bf x}_3$). We want to describe the scattering between a spin up particle and a dimer in the limite of zero incident energy. The total energy $E$ is therefore equal to the binding energy $E=-1/(2\mu\,a^2) <0$. In the limit where the spin down particle of position ${\bf x}_3$ is close to the spin up particle
of position ${\bf x}_2$, the wave function $\Psi$ is given by
\begin{eqnarray}
\Psi&\sim &\frac{f({\bf r})}{4\pi}\left(\frac{1}{r_{\perp}}-\sqrt{|E|}\right)\label{eqdeff}
\end{eqnarray}
where ${\bf r}_{\perp}=2\,\sqrt{m_{\uparrow}}({\bf x}_3-{\bf x}_2)$.
 In the limit ${\bf x}_2\to{\bf x}_3$, the vector
${\bf r}$ is given by ${\bf r}=\sqrt{m_{\uparrow}}({\bf x}_2-{\bf x}_1)/\sin \theta$ (therefore ${\bf r}$ has the dimension of a length multiplied by the square root of a mass, or
equivalently, of the inverse square root of an energy, since we take $\hbar\equiv 1$).
 The angle $\theta$ is given by $\tan \theta=\sqrt{1+2\,m_{\uparrow}/m_{\downarrow}}$.
In Ref.\cite{petr}, the following linear homogeneous equation for $f({\bf r})$ was derived 
\begin{eqnarray}
4\pi\int\,d{\bf r}'\,
\Big[
G_E(\sqrt{{\bf r}^2+{\bf r}'^2-2\,{\bf r}\cdot{\bf r}'\,\cos(2\theta)})f({\bf r}')
&&\nonumber\\
+\,
G_E(|{\bf r}-{\bf r}'|)\left(
f({\bf r})-f({\bf r}')\right)
\Big]&=&0\nonumber\\
\label{eqpetrovf}
\end{eqnarray}
with $G_E(|{\bf X}|)=\int\frac{d^6 {\bf K}}{(2\pi)^6}\,e^{i\,{\bf K}\cdot{\bf X}}
\frac{1}{\left(
|E|+{\bf K}^2\right)}$. 
We can Fourier transform Eq.(\ref{eqpetrovf}) and we find 
\begin{eqnarray}
\int\frac{d^3 {\bf q}'}{(2\pi)^3}
\frac{f({\bf q}')}{\left[|E|\,\sin^3 \theta + \left( 
{\bf q}^2 +{\bf q}'^2-2\,\cos \theta\; {\bf q}\cdot{\bf q}'
 \right)\,\sin \theta
 \right]}
&&\nonumber\\
+\,
\frac{1}{4\pi}\left[ \sqrt{|E|+{\bf q}^2}- \sqrt{|E|}\right]\,f({\bf q})
&=&0\nonumber\\
\label{eqftf}
\end{eqnarray}
Next, we define the function $g$ such that
\begin{eqnarray}
f({\bf q})&\equiv&(2\pi)^3\delta({\bf q})-\frac{4\pi}{{\bf q}^2}\gamma\,g(\gamma\,{\bf q})
\label{eqdefg}
\end{eqnarray} where $\gamma$ is a constant which is unknown {\it a priori}.
Inserting Eq.(\ref{eqdefg}) in Eq.(\ref{eqftf}), we find that we must take $\gamma =2\,a\,\sqrt{m_{\uparrow}}/\sqrt{1+2\,m_{\uparrow}/m_{\downarrow}}$ in order to recover Eq.(\ref{a_3}).
We thus have $g(\bar{{\bf q}})=\bar{a}_3(\bar{{\bf q}})$.

Integrating Eq.(\ref{eqdefg}) we can now express the integral relation Eq.(\ref{intrel}) for $\bar{a}_3$ in terms of  the function $f({\bf r})$ 
\begin{eqnarray}
\int \frac{d^3 {\bf q}}{(2\pi)^3}f({\bf q})&=&1-\int \frac{d^3 {\bf q}}{(2\pi)^3}
\frac{4\pi}{{\bf q}^2}\gamma\,g(\gamma\,{\bf q})\nonumber\\
=\,f({\bf r}=0)&=&1-\frac{2}{\pi}\int_0^{\infty}d\bar{q}\,\bar{a}_{3}(|\bar{q}|)\label{eqfaa3}
\end{eqnarray}
However, ${\bf r}=0$ corresponds to the situation where the two spin up particles are at the same position (${\bf x}_1 ={\bf x}_2$) and therefore the wave-function $\Psi$, hence $f$, must vanish due to the Pauli exclusion principle : $f({\bf r}=0)=0$.
From Eq.(\ref{eqfaa3}), we see that we must have
\begin{eqnarray}
\int_0^{\infty}d\bar{q}\,\bar{a}_{3}(|\bar{q}|)&=&\frac{\pi}{2}
\end{eqnarray}
which is precisely Eq.(\ref{intrel}). As a consequence, we see that the integral relation Eq.(\ref{intrel}) is simply equivalent to the statement that two spin up particles can not be at the same position in space. This also means that it is not expected 
to be correct for three identical bosons, for instance.

\end{document}